\documentclass[preprint,amsmath,amssymb,showpacs]{revtex4}

\usepackage{amssymb}
\usepackage{graphicx}
\usepackage{dcolumn}
\usepackage{amsmath}
\usepackage{subfigure}
\usepackage{bm}
\usepackage[latin1]{inputenc}
\newcommand{\comment}[1]{}

\begin{document}




\title{Depletion of nonlinearity in two-dimensional turbulence}


\author{Andrey V. Pushkarev and Wouter J.T. Bos}

\affiliation{LMFA, CNRS, Ecole Centrale de Lyon, Universit\'e de Lyon, 
  69134 Ecully, France}

\begin{abstract}
The strength of the nonlinearity is measured in decaying two-dimensional turbulence, by comparing its value to that found in a Gaussian field. It is shown how the nonlinearity drops following a two-step process. First a fast relaxation is observed on a timescale comparable to the time of formation of vortical structures, then at long times the nonlinearity relaxes further during the phase when the eddies merge to form the final dynamic state of decay. Both processes seem roughly independent of the value of the Reynolds number. 
\end{abstract}


\pacs{47.27.eb , 47.27.Gs, 47.27.Jv }
\maketitle


\section{Self-organization, coherence and depletion of nonlinearity}

Turbulent flows display the tendency to generate flow structures, starting from an unorganized initial state. This structuring of turbulent flows is perhaps most impressive in freely evolving two-dimensional turbulence, where large scale vortices are generated through nonlinear interaction \cite{McWilliams1984}. Even though intuitively one easily appreciates from visualisations that this self-organization of the flow increases its spatio-temporal coherence, it is not straightforward to give a quantitative measure of this coherence. In the present investigation we study the link between flow coherence and the strength of the nonlinearity during the self-organization process of two-dimensional turbulence starting from an initial state consisting of statistically independent modes. We will first illustrate how these quantities are related in nonlinear systems in general.

To illustrate our approach we consider a nonlinear system of the form
\begin{equation}\label{eq:1}
 \partial_t{\psi}=N(\psi^2)+ L(\psi),
\end{equation}
where $\psi(\bm x,t)$ is a function of space and time, $N(\psi^2)$ is the nonlinear term and $L(\psi)$ represents the linear terms. 
As a measure of the temporal coherence created by the self-organization process we can evaluate the rate at which on average the system evolves in time. Such a measure is  $\left<|\partial_t{\psi}|^2\right>$, where the brackets denote a space or ensemble average. If the mean-square time derivative decreases, it indicates that the temporal coherence of the considered field increases.

Squaring both sides of equation (\ref{eq:1})
\begin{equation}
\left<|\partial_t{\psi}|^2\right>=\left<|N(\psi^2)+ L(\psi)|^2\right>,
\end{equation}
we directly see that the coherence will be determined by an interplay of linear and nonlinear effects. A purely linear system will not display an interesting self-organization in the sense that the different modes are not coupled in the absence of nonlinearity. To simplify the problem we could therefore  consider a purely nonlinear system in the absence of linear terms. In that case we have 
\begin{equation}\label{eq:coh}
\left<|\partial_t{\psi}|^2\right>=\left<|N(\psi^2)|^2\right>.
\end{equation}
This shows that in the most elementary case of the nonlinear evolution of a system in the absence of any linear effects, the average temporal coherence is  completely determined by the mean-square nonlinearity. We will consider the evolution of the latter quantity in two-dimensional turbulent flows. It is well known for freely decaying two-dimensional turbulence in a periodic domain that the final dynamic state consists of two slowly evolving coherent vortices. The vorticity of the resulting flow-field shows a monotonic correlation with the stream-function \cite{Joyce1973}, corresponding to an equilibrium-state where the nonlinear interactions are small. Since coherent structures are also observed at shorter times, local equilibrium states should reduce the nonlocal interactions already at short and intermediate times and we expect thus the mean-square nonlinearity to be small compared to the quantity in a non-coherent velocity field even at these shorter times.  The definition of a non-coherent field 
is an interesting question 
itself. We will follow the idea of Kraichnan and Panda \cite{Kraichnan1988} and compare to a field with the same energy distribution, consisting of independent Fourier modes. This choice is not unique and does not allow to meaningfully quantify all possible non-Gaussian features of turbulence, as for instance enstrophy production \cite{Tsinober1999}. For the present purpose, which is the analysis of the depletion of the mean-square nonlinear term, it seems however a good choice. 

The mean-square nonlinearity as compared to its value in a Gaussian field was first investigated by Kraichnan and coworkers~\cite{Kraichnan1988,Chen1989,Kraichnan1989} in three dimensional decaying turbulence. Further studies \cite{Shtilman1989,Tsinober1990,Tsinober1999,Tsinober} aimed at indentifying mechanisms which can explain the depletion of nonlinearity through local flow structuring. One such mechanism, the alignment of velocity and vorticity, was suggested to play an important role in the dynamics of turbulence \cite{Pelz1985}. The study of Kraichnan and Panda \cite{Kraichnan1988}, put forward the idea that velocity-vorticity alignment might be one manisfestation of a more generic feature of turbulent flows, which is its tendency to diminish the strength of the nonlinearity. This seems indeed to be the case since the depletion of nonlinearity is also observed in magnetohydrodynamic turbulence \cite{Servidio2008} and the advection term in the advection of a passive scalar also shows this tendency 
\cite{Herring1992,Bos2012-3}.

Our study is a logical extension of these works to the case of two-dimensional turbulence. One first lesson that we can learn from the outcome of the present work is that, if the nonlinearity is reduced in two-dimensional turbulence, it is not the velocity-vorticity alignment which is the underlying mechanism which explains the depletion of nonlinearity in both two- and three-dimensional turbulence, since the vorticity is always perpendicular to the velocity in two-dimensional flow. Alignment, in general, can however play an important role, as was shown in \cite{Servidio2010}, where a preferential alignment of the vorticity-gradient and stream-function gradient was observed in decaying two-dimensional turbulence.

With respect to this latter work, the originality of the present approach is that we will compare statistics in a decaying turbulent flow with a Gaussian reference field having the same energy spectrum at every time-instant, as proposed  in \cite{Kraichnan1988}. It allows to show how the temporal coherence of the flow evolves in time. We do not focus on the precise local relaxation mechanism and argue, following \cite{Kraichnan1988}, that it is perhaps not the detailed topological mechanism which is universal, but the tendency to create coherence by decreasing, statistically, the norm of the nonlinear term.


\section{The depletion of nonlinearity in two-dimensional turbulence}

For an incompressible fluid flow, equation (\ref{eq:1}) is given by the Navier-Stokes equations 
\begin{eqnarray}
 \partial_t \bm u=-\bm N+\nu\Delta \bm u, ~~~~~\nabla\cdot \bm u=0,
\end{eqnarray}
with the nonlinear term given by
\begin{eqnarray}
\bm N=\bm u\cdot \nabla \bm u+\nabla p,
\end{eqnarray}
with $p$ the pressure divided by the fluid density, which is assumed homogeneous and constant, and $\nu$ the kinematic viscosity. In two-dimensions it is convenient to consider the evolution of $\omega=\bm e_\perp \cdot \nabla\times \bm u$, the component of the vorticity normal to the plane of the velocity,
\begin{eqnarray}\label{eq:vortEq_2D}
 \partial_t \omega=-N_\omega+\nu\Delta \omega,
\end{eqnarray}
with
\begin{eqnarray}
N_\omega&=&\bm e_\perp \cdot \nabla\times \bm N\nonumber\\
&=& \bm u \cdot \nabla \omega.
\end{eqnarray}
In the absence of linear effects, it will be the quantity $\langle N_\omega^2 \rangle$, which measures the temporal Eulerian coherence of the vorticity field, as suggested by equation (\ref{eq:coh}). For a strongly coherent field this quantity will be small compared to the quantity in a non-coherent velocity field. As explained above, we will compare to a field with the same energy distribution, consisting of independent Fourier modes, by measuring 
\begin{eqnarray}\label{eq:2D_N_over_Ng}
\frac{\langle N_\omega^{2}\rangle}{\langle N_{_\omega,G}^{2}\rangle},
\end{eqnarray}
where $\langle N_{_\omega,G}^{2}\rangle$ is the mean-square vorticity advection term measured in a multivariate Gaussian field with the same energy spectrum. The mean-square nonlinear term is not invariant under Galilean transformations, \emph{i.e.} adding a spatially uniform velocity field to the flow will change its value. Adding such a field will also change the estimate in the Gaussian field so that measure (\ref{eq:2D_N_over_Ng}) compensates, at least in part, for such effects. In the present work we will set the mean flow $\langle \bm u \rangle=0$ and only consider the statistically isotropic case.

The evolution of equation~(\ref{eq:vortEq_2D}) was computed using a two dimensional pseudo-spectral Fourier code in stream function-vorticity formulation. The computation was performed on a $(2\pi)^{2}$ square domain with periodic boundary conditions. The resolution was varied between $256$ and $2048$ Fourier modes in each direction. The initial velocity field is defined in Fourier space and corresponds to a field of incompressible statistically independent Fourier modes, given by
\begin{eqnarray}\label{eq:initVelocFld2D}
u_{i}(\bm k) = \sqrt{\frac{E(k)}{\pi k}}P_{ij}(\bm k)a_{j}(\bm k),
\end{eqnarray}
where $a_{i}(\bm k)$ is a Gaussian white noise process and $E(k)\sim k^{4}\exp(-k^{2}/k_{0}^{2})$ is the kinetic energy spectrum. The value of $k_0$ sets the dominant correlation wavenumber of the initial energy distribution and is $k_0=10$. The Riesz-operator, 
\begin{equation}
P_{ij}(\bm k)=\delta_{ij}-k_ik_j/k^2, 
\end{equation}
projects the random noise on the plane perpendicular to the wavevector to ensure incompressibility of the initial velocity field. The kinematic viscosity $\nu$ and the corresponding values for the initial Reynolds numbers based on the box-size and on the Taylor lengthscale, respectively, are summarized in table \ref{tab:1}.

\begin{table}
\begin{tabular}{c|| c| c| c| c}
\hline
\hline
Resolution          &  $256^2$ & $512^2$  & $1024^2$ & $2048^2$ \\
$\nu$ &$5.~10^{-3}$ &$2.~10^{-3}$ &$5.~10^{-4}$ &$2.~10^{-4}$\\
$R_L$ &$1256$ &$3141$ &$12566$ &$31415$\\
$R_\lambda$ &$40$ &$100$ &$400$ &$1000$\\
\hline
\hline
\end{tabular}
\caption{\label{tab:1} Details of the simulations. We define $R_L=UL/\nu$, with $U$ the rms-value of one of the velocity components and $L$ the box-size. The Taylorscale Reynolds number is defined as $R_\lambda=U\lambda/\nu$, with $\lambda=U\sqrt{15\nu/\epsilon}$.  }
\end{table}


\begin{figure}[]
\center{
\subfigure[]{\includegraphics[width=0.5\columnwidth,angle=0]{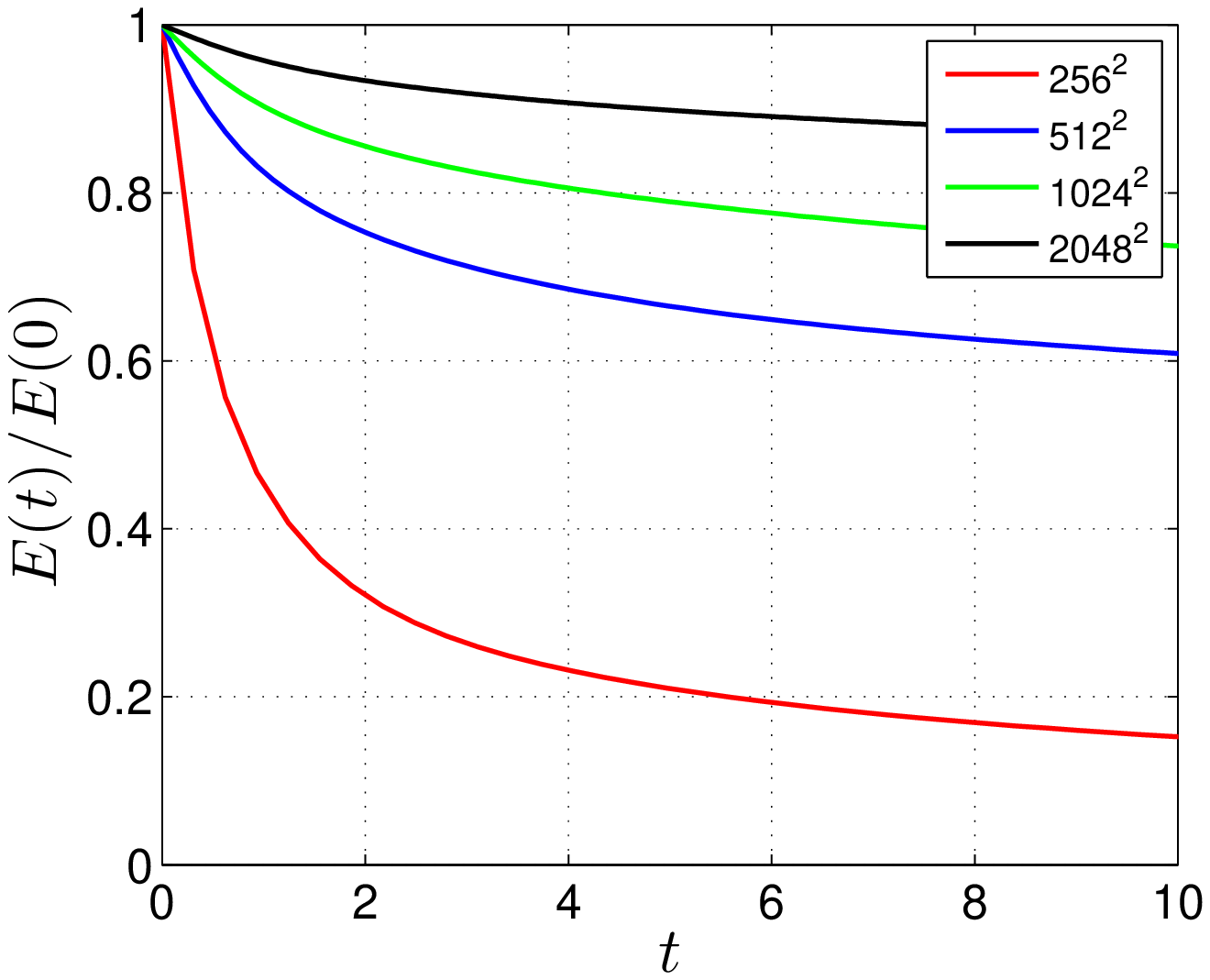}}~
\subfigure[]{\includegraphics[width=0.5\columnwidth,angle=0]{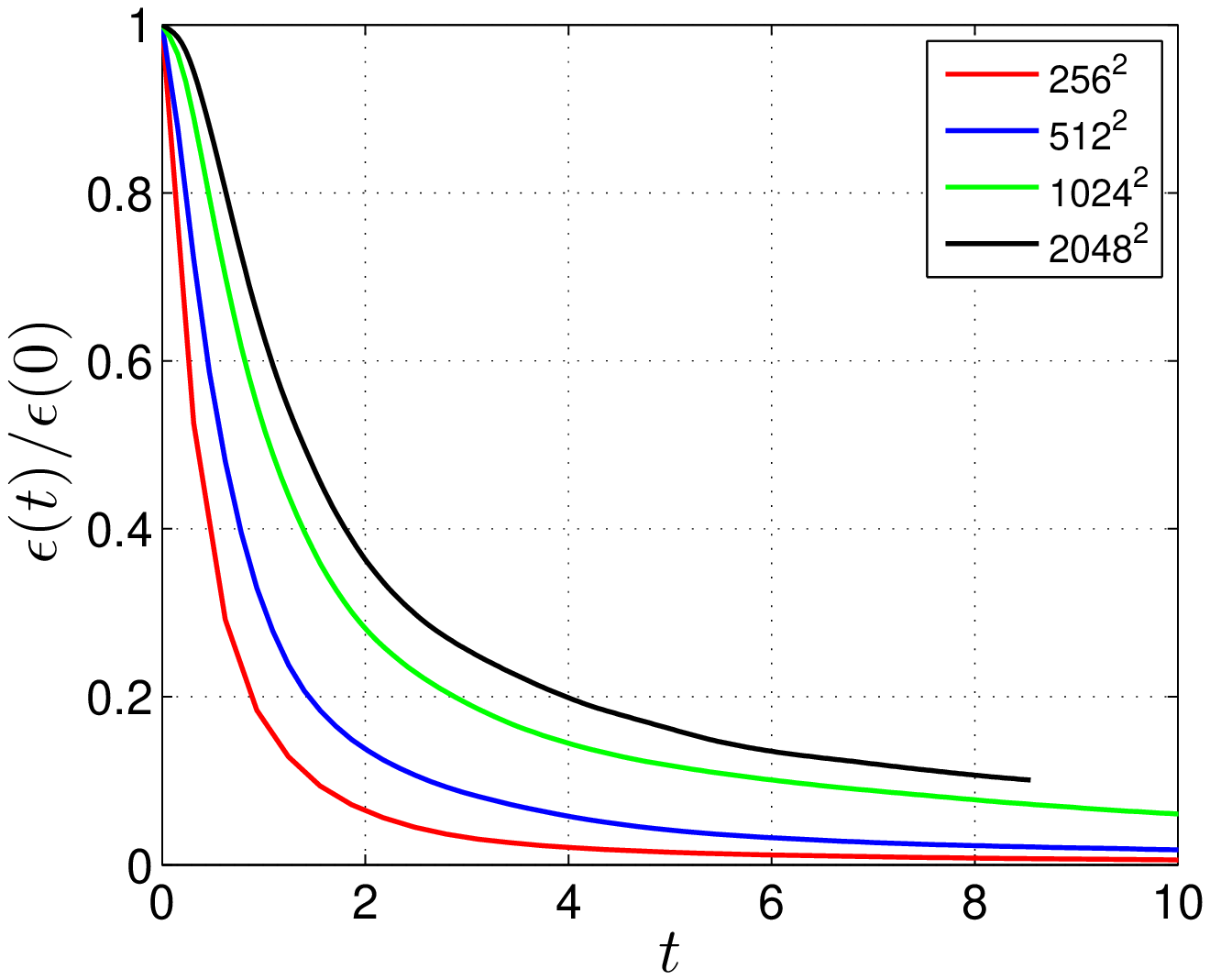}}}
\caption{Time evolution of the energy (a) and energy dissipation (b).}
\label{ris:Et}
\end{figure}

The temporal evolution of the kinetic energy $E=\left<u_iu_i\right>/2$ and enstrophy $Z=\left<\omega^2\right>/2$ is given by
\begin{eqnarray}
\partial_t E=-\epsilon\nonumber\\
\partial_t Z=-\eta,
\end{eqnarray}
with $\epsilon$ and $\eta$ the energy and enstrophy dissipation rate. In a periodic two-dimensional decaying flow without energy and enstrophy sources, both the kinetic energy and enstrophy are monotonically decaying functions of time. Since the energy dissipation $\epsilon=2\nu Z$, as soon as the enstrophy has decayed considerably, the energy will remain approximately constant at high Reynolds numbers. This behaviour is shown in figure \ref{ris:Et}, which illustrates the time evolution of the kinetic energy and energy dissipation rate, respectively. 

\begin{figure}[]
\center{
\subfigure[]{\includegraphics[width=.5\textwidth,angle=0]{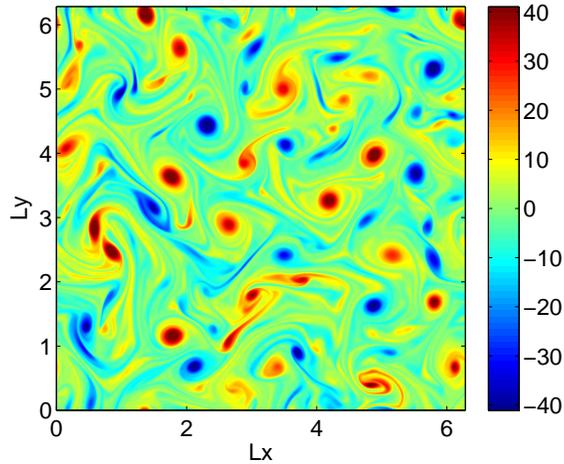}}
\subfigure[]{\includegraphics[width=.5\textwidth,angle=0]{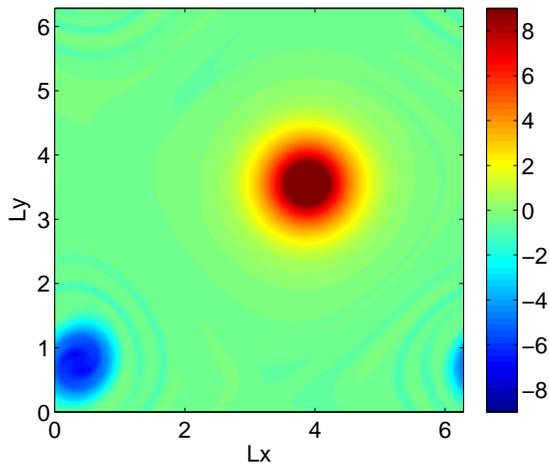}}
}
\caption{Snapshots of the vorticity field at $t\approx3$ (a) and at $t\approx85$ (b).}
\label{ris:snapshots_vortFld_2D}
\end{figure}

Figures~\ref{ris:snapshots_vortFld_2D} depict snapshots of the vorticity fields of the $1024^2$ run at $t\approx3$ (a) and at $t\approx85$ (b). The condensation of all the vorticity into two counterrotating vortices is observed for long times. It will be shown now that this state corresponds to a flow with a very small magnitude of the nonlinearity.

\begin{figure}[]
\center{
\subfigure[]{\includegraphics[width=.5\textwidth,angle=0]{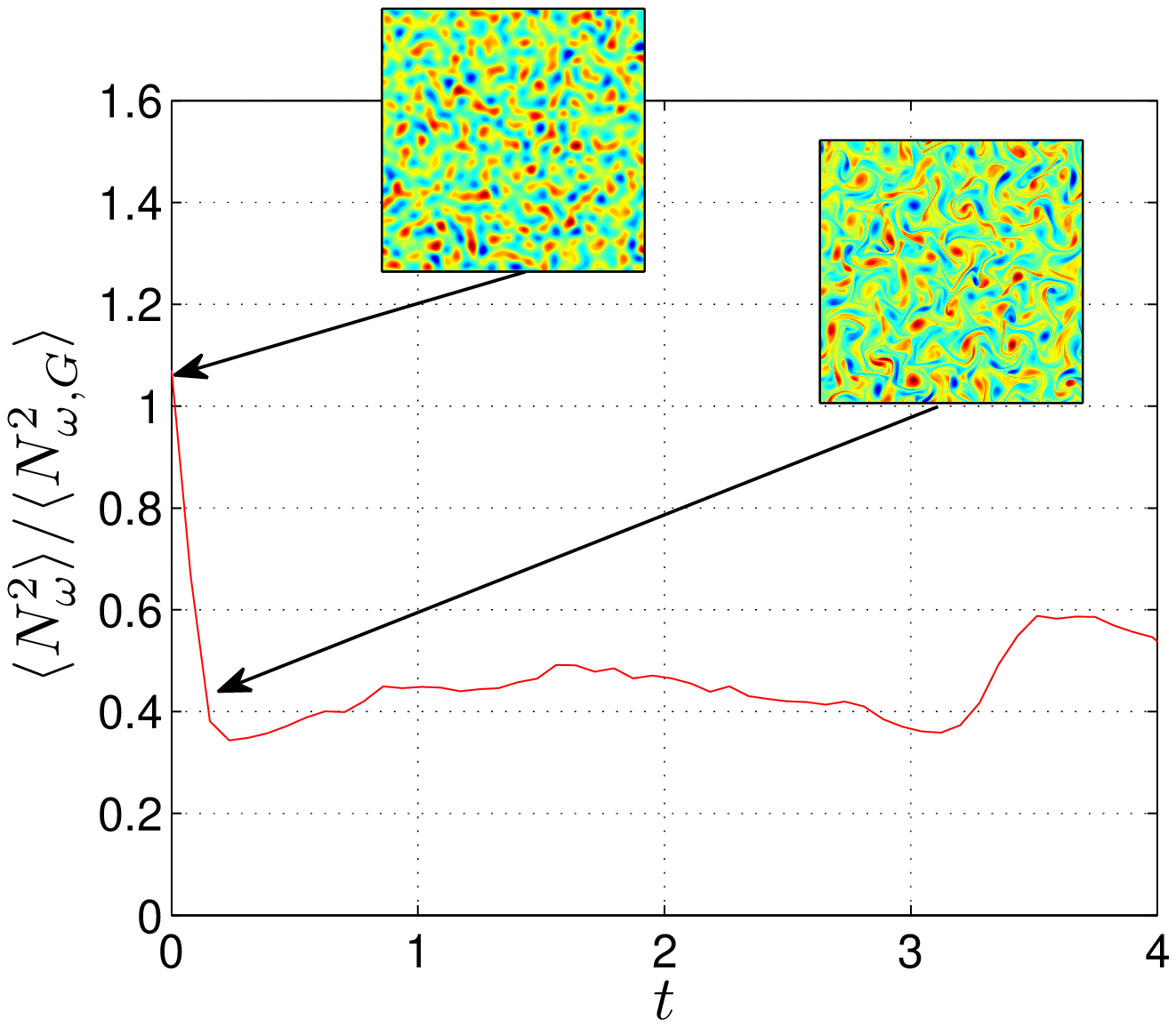}}
\subfigure[]{\includegraphics[width=.5\textwidth,angle=0]{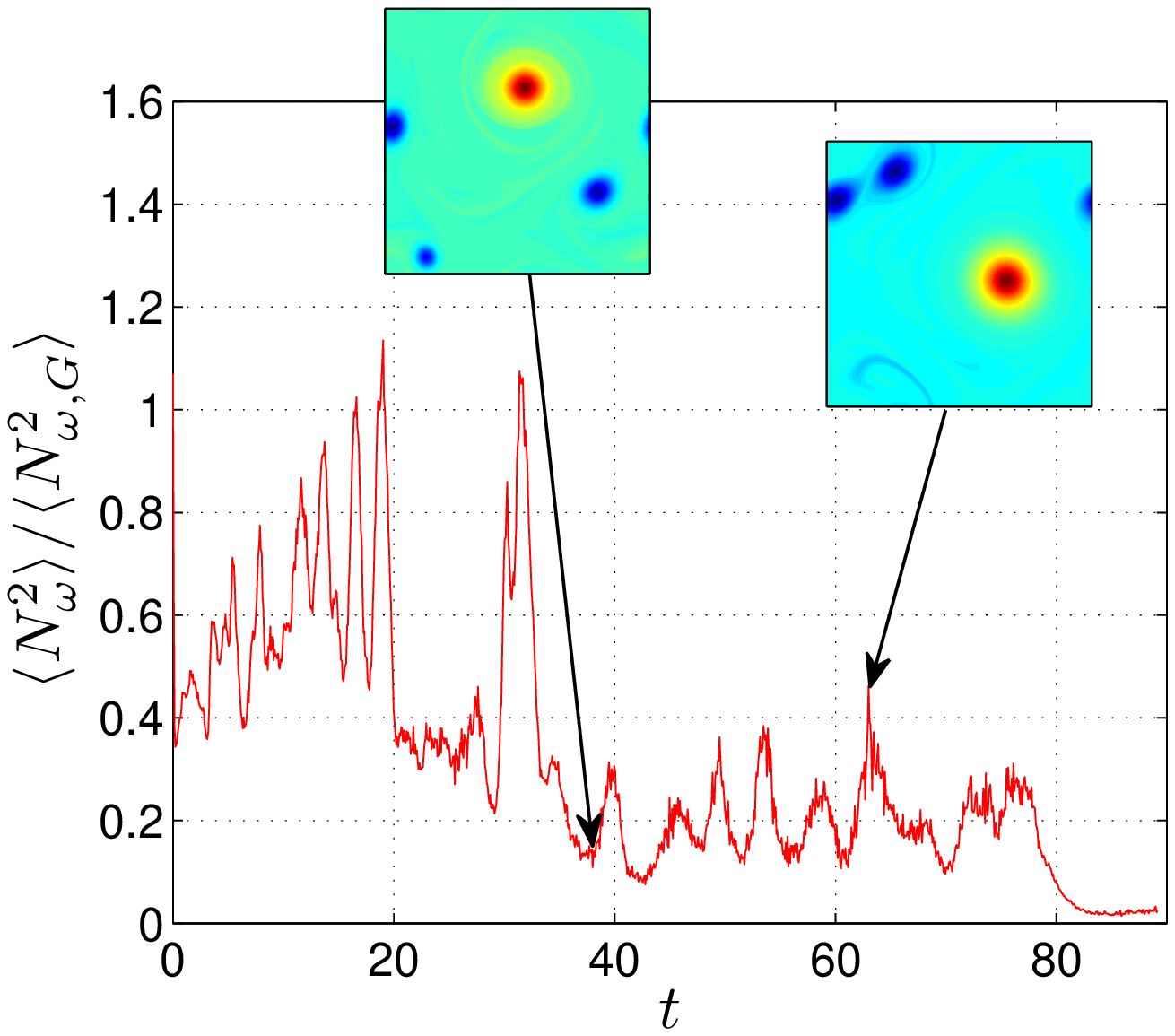}}}
\caption{Evolution of the ratio of the total mean square nonlinear term in turbulent flow to the equivalent value in the Gaussian flow for short times (a) and long times (b). Snapshots of the vorticity field are shown for four different time instants. In the top figure these snapshots only show a quarter of the domain to more clearly identify the flow structures. Results correspond to the $1024^2$ run.}
\label{ris:N_over_Ngaus_snapshots}
\end{figure}

In figure~\ref{ris:N_over_Ngaus_snapshots} we show the evolution of $\langle N_\omega^2\rangle/\langle N_{\omega,G}^2\rangle$.
As we can see from the figure, this ratio very rapidly decreases at the initial time. After this moment, during the further evolution of the flow, peaks are observed which correspond to the interaction between different vortices. The opposite, local minima correspond to a quiet state where there is no interaction between different vortices and all vortices are well separated. This is illustrated by the snapshots which are present on the figure. The moments of time which correspond to the snapshots are noted by arrows. We have checked that the same qualitative picture is observed during the other peaks and local minima of the time-evolution. The fact that vortex merging increases the value of the nonlinearity can be understood intuitively since when two vortices approach, the relative velocity between the cores of the vortices is parallel to their radial vorticity gradients so that the value of $(\bm u\cdot \nabla \omega)^2$ is high during these events. 

\begin{figure}[]
\center{
\includegraphics[width=.5\textwidth,angle=0]{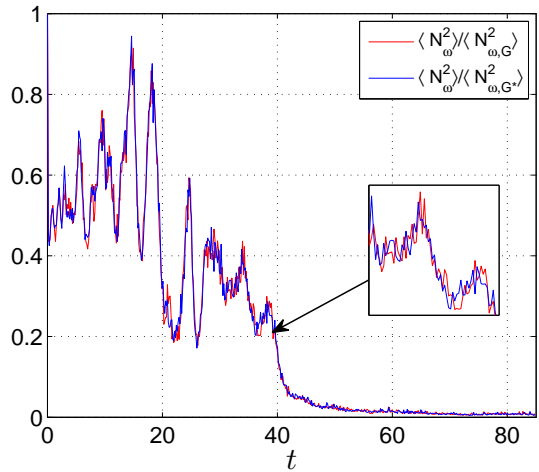}}
\caption{Evolution of the ratio of the total mean square nonlinear term in turbulent flow to the equivalent value in the Gaussian flow. The turbulent statistics for the $512^2$ run are compared to two independently generated Gaussian fields $\langle N_{\omega,G}^2\rangle$ and $\langle N_{\omega,G^*}^2\rangle$ to check the influence of the choice of the Gaussian fields on the results.}
\label{ris:Check}
\end{figure}

We have checked that the order unity peaks are not an artefact of the Gaussian random fields we used for comparison. In figure \ref{ris:Check} we show the results for the mean-square nonlinearity compared to two independently generated Gaussian fields for the $512^2$ run. It is observed that the low frequency structure of the time evolution is not significantly affected when changing the reference field. Only a zoom at the fine time-structure shows a difference in behaviour.

At the end of the simulation only two weakly interacting vortices are observed, figure~\ref{ris:snapshots_vortFld_2D}~(b). At this time it is observed that the mean square nonlinearity is reduced to only a few percent of its Gaussian value.

It seems that in the two-dimensional case there are two timescales over which the nonlinearity is reduced: a very short one at the beginning of the simulation of the order of time over which the initial vortices form, and a long one, of the order of the time it takes to condense all the energy in the two counterrotating vortices in figure~\ref{ris:snapshots_vortFld_2D}. 

\begin{figure}[]
\center{
\includegraphics[width=8.5cm,angle=0]{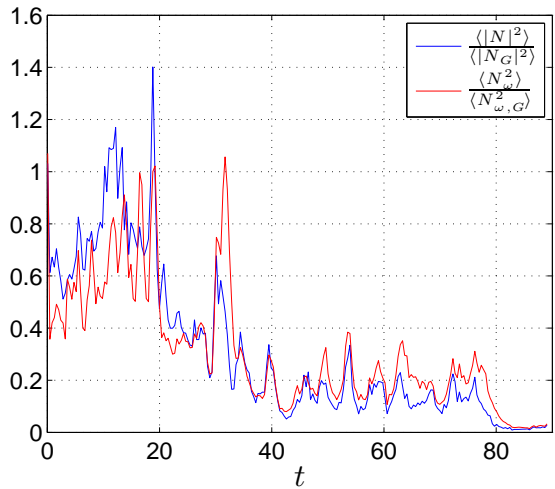}}
\caption{Time evolution of the total mean square nonlinear term in turbulent flow to the equivalent value in the Gaussian flow for the values $\langle |\bm N|^2\rangle/\langle |\bm N_G|^2\rangle$ and $\langle N_\omega^2\rangle/\langle N_{\omega,G}^2\rangle$.}
\label{ris:twoDifN}
\end{figure}

To give a statistical description of the multiscale character of the depletion of nonlinearity in 2D turbulence we define the nonlinearity spectrum of the Navier-Stokes equations as
\begin{eqnarray}
\int W(k)dk= \langle |\bm N|^2 \rangle.
\end{eqnarray}
The vorticity nonlinearity spectrum is given by
\begin{eqnarray}
\int W_\omega(k)dk= \langle N_\omega^2 \rangle
\end{eqnarray}
and the relation between the two spectra is
\begin{eqnarray}
W_\omega(k)= k^2W(k).
\end{eqnarray}
Therefore, the ratio of the spectrum of the mean square nonlinear term in the turbulent flow to the equivalent value in the Gaussian flow does not depend on which nonlinearity we consider:
\begin{eqnarray}
\frac{W_\omega(k)}{W_{\omega,G}(k)}= \frac{W(k)}{W_{G}(k)}.
\end{eqnarray}
However, the ratios of the integrated quantities are not necessarily the same, which is illustrated in figure~\ref{ris:twoDifN}, where we compare $\langle |\bm N|^2\rangle/\langle |\bm N_G|^2\rangle$ and $\langle N_\omega^2\rangle/\langle N_{\omega,G}^2\rangle$ for the $1024^2$ run. It is shown, however, that the two ratios are very close.

\begin{figure}[]
\center{
\subfigure[]{\includegraphics[width=.5\textwidth,angle=0]{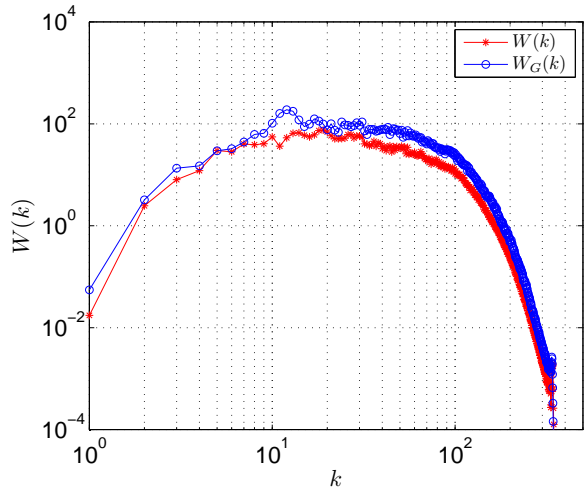}}
\subfigure[]{\includegraphics[width=.5\textwidth,angle=0]{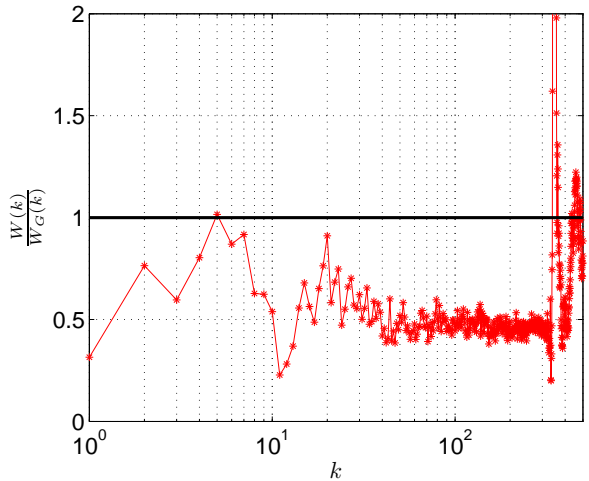}}}
\caption{Top: spectrum of the nonlinear term $W(k)$ and $W_G(k)$ evaluated at $t=4$. Bottom: normalized power spectrum $W(k)/W_G(k)$ of the nonlinear term evaluated at $t=4$.}
\label{ris:specWk_Norm_2D}
\end{figure}

We show $W(k)$ and $W_G(k)$ at $t=4$ in figure~\ref{ris:specWk_Norm_2D} (a). It is shown that the spectra are an approximately constant function of the wavenumber in the intermediate range of wavenumbers. The ratio $W(k)/W_G(k)$ is shown in figure~\ref{ris:specWk_Norm_2D} (b).
It is shown that the depletion takes place in almost the complete spectrum. In three dimensional turbulence, according to closure theory, the depletion was constant throughout the inertial range of scales \cite{Bos2013-2}. Note that the highest wavenumbers, beyond $k=300$, are affected by the wavenumber truncation and the strong fluctuations at these scales are rather numerical artefacts than part of the turbulent dynamics.

\begin{figure}[]
\center{
\subfigure[]{\includegraphics[width=.5\textwidth,angle=0]{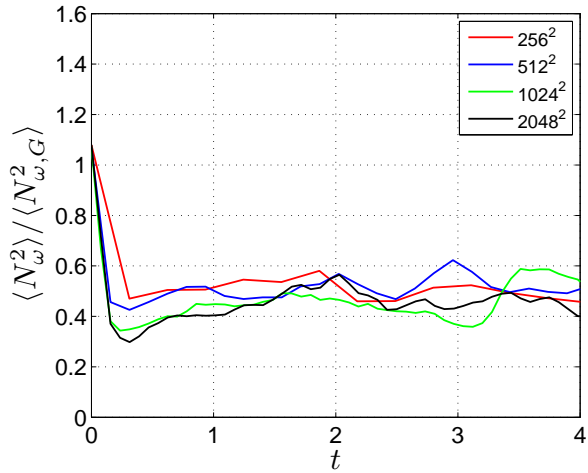}}
\subfigure[]{\includegraphics[width=.5\textwidth,angle=0]{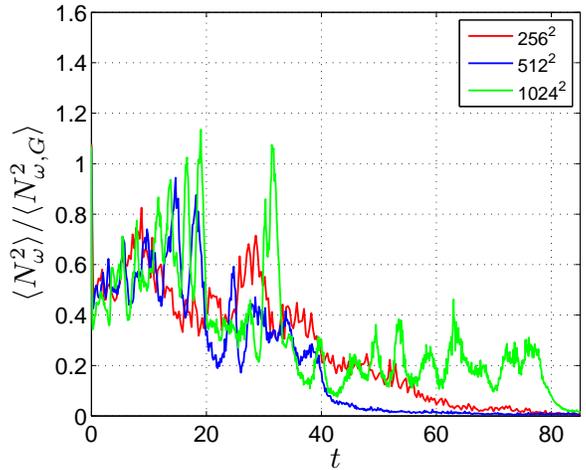}}}
\caption{Evolution of the ratio of the total mean square nonlinear term in turbulent flow to the equivalent value in the Gaussian flow for short times (a) and long times (b).}
\label{fig:Reeff}
\end{figure}

An obvious question is of course how the above results depend on the initial value of the Reynolds number. Are the results essentially inviscid, so that we can expect them to hold in the limit of infinitely high Reynolds numbers, or do they depend on the relative strength of the viscous effects compared to that of the nonlinear term? In figure \ref{fig:Reeff} the results are shown for the different simulations. It is observed that the rapid relaxation becomes roughly independent of the Reynolds number for the $512^2$, $1024^2$ and $2048^2$ simulations. The long-time evolutions are qualitatively similar. For numerical reasons, we have not carried out the $2048^2$ simulation for long times, but the qualitative similarity between the other three simulations indicates that the long-time evolution, like the short-time evolution, is not qualitatively changed by the value of the Reynolds number. Both long and short time processes appear to be inviscid.

\section{Discussion and conclusion}

In the present investigation we have shown how the mean-square nonlinearity evolves in 2D turbulence, compared to a Gaussian reference field. In particular, we observed a two-step regime of the suppression of nonlinearity. The first step corresponds to the fast formation of vortices and could be qualified as rapid relaxation. The second phase corresponds to the evolution of the system through the mutual interactions of the vortices. If we consider only a short time interval, as in figure~\ref{ris:N_over_Ngaus_snapshots} (Top), it seems that the mean square nonlinearity falls to a value of the order of half the Gaussian value and then remains constant, as seen qualitatively in  3D turbulence (e.g. \cite{Kraichnan1988,Bos2013-2}). Comparing the strongly fluctuating time-evolution of the ratio $\langle N_\omega^{2}\rangle/\langle N_{_\omega,G}^{2}\rangle$ with flow-visualizations at different time-instants, it was observed that low values correspond to situations where vortical structures are well 
separated in space. Whenever vortices interact, a strong increase of the ratio is observed. 

The first, rapid relaxation process is possibly caused by local alignment properties of the flow \cite{Servidio2010}. The fact that at very long times the nonlinearity is further reduced was predicted by an entropy optimization argument by Joyce \& Montgomery~\cite{Joyce1973}, and more formally explained in \cite{Robert1991,Miller1990}. Note that this final state of two vortices, depleted from nonlinear interaction was predicted by applying statistical mechanics to the global quantities (enstrophy, energy averaged over the spatial domain) leading to the prediction of the most probable final dynamic state in the limit of small viscosity.  A recent review article  \cite{Bouchet2012} gives an overview of existing attempts to predict the formation of coherence using the application of statistical mechanics to two-dimensional flows. These approaches mostly focus on the long-time limit of systems in the inviscid limit. The results in the present work suggest that also the rapid relaxation process is of 
inviscid nature, and might therefore also be treatable in the framework of Euler's equations.

We have not investigated the role of the boundary conditions on the depletion of nonlinearity. It is certain that the detailed long-time dynamics must be affected by this. Indeed, it was shown that the final dynamic state is different when an unbounded fluid is considered \cite{Gallay2005,Montgomery2011} or when rigid boundaries are taken into account \cite{Pointin1976,Taylor2009}. However, in all these cases the final state is characterized by large-scale coherent structures, that we can expect to be in a near equilibrium state with a low value of the mean-square nonlinearity, compared to a field consisting of random independent modes. The long-time tendency of the mean-square nonlinearity will thus not be changed imposing different boundary conditions. Furthermore, it can be expected that the initial fast relaxation is relatively independent of the boundary conditions, as long as the initial correlation lengthscale is small compared to the domain-size.

Transposing the existing theoretical ideas of inviscid global relaxation  to local flow structuring, might prove a good starting point to understand the rapid relaxation events observed in the present and different flow configurations. A suggestion of such an approach and first steps in this direction can be found in the work by Servidio {\it et al.} \cite{Servidio2010}. Such an approach could lead the way to a statistical 
understanding of the self-organization of turbulent systems in general without appeal to specific topological considerations of the local flow structures.

\section*{Acknowledgements}

The authors are grateful to Robert Rubinstein for comments, valuable discussions and suggestions.


\end{document}